\documentclass[manuscript,screen]{acmart}

\AtBeginDocument{%
  }

\usepackage{enumitem}
\usepackage{xcolor}
\usepackage{tabularx}
\usepackage{longtable}
\usepackage{ragged2e}
\usepackage{multirow}
\usepackage{listings}
\usepackage{arydshln}

\usepackage{float}
\newfloat{metric}{htbp}{lom}[section]
\floatname{metric}{Metric}

\newif\ifdraft
 \draftfalse

\ifdraft
  \newcommand{\todo}[1]{\textsf{\textbf{\textcolor{red}{[TODO: #1]}}}}
\else
  \newcommand{\todo}[1]{}
\fi

\ifdraft
  \newcommand{\rev}[1]{\textcolor{blue}{#1}}
\else
  \newcommand{\rev}[1]{#1}
\fi

\begin{document}


\title{Empowering Affected Individuals to Shape AI Fairness Assessments: Processes, Criteria, and Tools}

\author{Lin Luo}
\authornote{Corresponding Author}
\orcid{https://orcid.org/0000-0002-0310-3158}
\email{l.luo.1@research.gla.ac.uk}
\affiliation{%
  \institution{School of Computing Science, University of Glasgow}
  \country{Glasgow, United Kingdom}
}

\author{Satwik Ghanta}
\orcid{https://orcid.org/0009-0003-1110-055X}
\email{y.ghanta.1@research.gla.ac.uk}
\affiliation{%
  \institution{School of Computing Science, University of Glasgow}
  \country{Glasgow, United Kingdom}
}

\author{Yuri Nakao}
\orcid{https://orcid.org/0000-0002-6813-9952}
\email{nakao.yuri@fujitsu.com}
\affiliation{%
  \institution{FUJITSU LIMITED}
  \country{Tokyo, Japan}
}
\affiliation{%
  \institution{Graduate School of Arts and Sciences, The University of Tokyo}
  \country{Tokyo, Japan}
}

\author{Mathieu Chollet} 
\orcid{https://orcid.org/0000-0001-9858-6844}
\email{mathieu.chollet@glasgow.ac.uk}
\affiliation{%
  \institution{School of Computing Science, University of Glasgow}
  \country{Glasgow, United Kingdom}
}

\author{Simone Stumpf}
\orcid{https://orcid.org/0000-0001-6482-1973}
\email{simone.stumpf@glasgow.ac.uk}
\affiliation{%
  \institution{School of Computing Science, University of Glasgow}
  \country{Glasgow, United Kingdom}
}

\renewcommand{\shortauthors}{Luo et al.}

\begin{abstract}
AI systems are increasingly used in high-stakes domains such as credit rating, where fairness concerns are critical. Existing fairness assessments are typically conducted by AI experts or regulators using predefined protected attributes and metrics, which often fail to capture the diversity and nuance of fairness notions held by the individuals who are affected by these systems' decisions, such as decision subjects. Recent work has therefore called for involving affected individuals in fairness assessment, yet little empirical evidence exists on how they create their own fairness criteria or what kinds of criteria they produce - knowledge that could not only inform experts' fairness evaluation and mitigation, but also guide the design of AI assessment tools. We address this gap through a qualitative user study with 18 participants in a credit rating scenario. Participants first articulated their fairness notions in their own words. Then, participants turned them into concrete quantified and operationalized fairness criteria, through an interactive prototype we designed. Our findings provide empirical evidence of the process through which people's fairness notions emerge via grounding in model features, and uncover a diverse set of individuals' custom-defined criteria for both outcome and procedural fairness. We provide design implications for processes and tools that support more inclusive and value-sensitive AI fairness assessment. 

\end{abstract}



\maketitle

\section{Introduction}
Artificial Intelligence (AI) systems increasingly govern high-stakes decisions like loan approvals and job offers, yet many exhibit unfairness \cite{10.1145/3715275.3732004, genderbiasincredits, COMPAS}. Currently, AI experts (i.e., AI fairness researchers and practitioners) or external regulators (e.g., legal \& compliance teams) \cite{landers2023auditing, 10.1145/3593013.3594076} have assessed AI fairness, choosing their own criteria \cite{FairnessAssessmentPractionerPerspective}. Typically, this involves selecting protected features (e.g., gender and race), and using pre-defined fairness metrics and accepted thresholds to assess fairness or discrimination with respect to those features \cite{study3, saleiro2018aequitas}. 

However, recent efforts emphasize involving broader stakeholder types in fairness assessment to promote more inclusive and publicly supported AI systems, especially \textit{affected individuals}, those who are subject to AI-driven decisions but lack AI expertise, i.e., formal AI training or professional experience \cite{10.1145/3631700.3664912,2024crowdsource/10.1145/3640543.3645209,10.1145/3630106.3659044, 10.1145/3359283,10.1145/3514258,earnfairness}. Our work focuses on such decision subjects from the general public (hereafter \textit{stakeholders}, excluding AI experts and external regulators). This inclusion has been motivated by three main reasons. First, fairness is a human value with diverse notions difficult to express as a single metric \cite{10.1145/3711079,FairnessAssessmentPractionerPerspective}. Second, existing metrics often cannot capture fairness nuances and require stakeholder input \cite{FairnessAssessmentPractionerPerspective, 10.1145/3711079}. Third, it has been shown that stakeholders frequently hold fairness notions that diverge from the current prevalent practices of AI experts \cite{study3,earnfairness,10.1145/3411764.3445308}. 

Existing approaches typically involve stakeholders in assessment by directly asking them for their metric preferences selected from commonly used AI expert metrics \cite{earnfairness}, or indirectly eliciting these preferences through choosing between alternative models or prediction outcomes~\cite{2024crowdsource/10.1145/3640543.3645209,10.1145/3411764.3445308, Saxena/10.1145/3306618.3314248,Yokota2022Toward,MathematicalNotions/10.1145/3292500.3330664}. Recent work has begun eliciting stakeholders' custom fairness notions in their own language \cite{study3,10.1145/3411764.3445308,10.1145/3630106.3659044}. However, existing work does not yet empower stakeholders to articulate fairness and turn these values into quantified, operational fairness criteria, risking misinterpretation of stakeholder values \cite{manders2011values}. We face a knowledge gap in how stakeholders form fairness notions, turn them into quantifiable, operationalizable criteria, and what types of criteria result. Addressing this gap could inform fairness assessment tool design and AI experts' criteria selection for assessment and mitigation \cite{hort2023bias, 8843908, de2025towards}.
 
We address this gap by uncovering how stakeholders create fairness criteria and what they produce. We conducted a user study with 18 participants in a credit rating scenario. Participants first reflected on their fairness notions in their own lay language. They then turned these notions into concrete criteria that were able to be quantified and operationalized for assessing fairness with the support of an interactive prototype we designed and researcher facilitation, which scaffolded rather than prescribed, criteria creation. The prototype enabled participants to translate their fairness notions into operational criteria by using existing outcome metrics, combining existing metrics, or creating entirely new metrics, and to apply these criteria in real time to assess AI fairness. Through this study, we addressed the following research questions:

\begin{itemize}
    \item \textbf{RQ1:} How do stakeholders create their own fairness criteria?

    \item \textbf{RQ2:} What kinds of fairness criteria do stakeholders create?
\end{itemize}

Our work makes three main contributions. First, we uncover how stakeholders create fairness criteria, tracing the process from their initial plain-language fairness notions to the development of concrete criteria that can be operationalized in assessing fairness. Second, we provide empirical insights into the kinds of fairness criteria that stakeholders create. Third, we make suggestions to design processes and tools to support stakeholders to create fairness criteria based on their own fairness values and notions, thereby paving the way for more inclusive and responsible AI fairness assessment and, ultimately, AI systems.


\section{Related Work}
\subsection{Assessing AI Fairness: Current Practice and Its Limitations}
\label{sec: AI fairness assessment practice}
AI experts or regulators \cite{landers2023auditing} have focused on \textit{outcome fairness}, assessing whether AI systems' outcomes are fair \cite{hort2023bias, 10.1145/3457607, 10.1145/3494672, FairnessAssessmentPractionerPerspective}. This typically starts by selecting protected demographic attributes \cite{hort2023bias,10.1145/3457607,10.1145/3194770.3194776,FairnessAssessmentPractionerPerspective, 10.1145/3715275.3732040, 10.1145/3313831.3376445}, including legally protected feature such as gender and race \cite{EqualityAct2010,EUCharter2012}, their intersections \cite{FaccTSubgroupFairness10.1145/3287560.3287592}, as well as \textit{sensitive features} that may indirectly encode them e.g., foreign worker status might indirectly reflect race \cite{FairnessDatasetSurvey}. Second, fairness metrics are chosen to quantify fairness. Existing fairness metrics \cite{10.1145/3194770.3194776} are commonly divided into two categories \cite{conflict/10.1145/3351095.3372864}. \textit{Group fairness metrics} examine whether different groups are treated equally by the system by comparing \textit{unprivileged/protected groups} versus \textit{privileged/unprotected groups} (e.g., gender: female versus male) \cite{10.1145/3457607}. Demographic Parity (DP), one of the most popular metrics, requires that different groups receive favorable outcomes at the same rate, regardless of other factors \cite{10.1145/2090236.2090255, NIPS2017_a486cd07}. \textit{Individual fairness metrics} focus on whether similar individuals are treated similarly, such as Consistency metric \cite{pmlr-v28-zemel13}. In practice, group fairness is easier to quantify statistically, while individual fairness is harder to operationalize because defining ``similarity'' between individuals is ambiguous and difficult to measure in a universally accepted way \cite{conflict/10.1145/3351095.3372864}. Metric selection lacks clear guidelines and is often driven by legal requirements (e.g., Demographic Parity) or established organizational practices \cite{FairnessAssessmentPractionerPerspective,hort2023bias, 10.1145/3457607}. A common practice is to choose one metric \cite{hort2023bias} and use it across multiple (protected) features \cite{10.1145/3442188.3445902, 10.1145/3531146.3533225, 10.1145/2783258.2783311}. Finally, what counts as fair is determined by setting \textit{fairness thresholds}, which define how much unfairness is acceptable. These thresholds lack a unified standard; a common practice is to treat the Demographic Parity Ratio (DPR) above 0.8 as fair under the law \cite{10.1145/2783258.2783311}.

Increasing attention has also been given to \textit{procedural fairness} \cite{Robert2020Designing,10.1145/3359284}, which focuses on whether AI's decision processes are fair~\cite{10.1145/3359284}. Procedural fairness is more difficult to assess than outcome fairness, so standardized or unified metrics have not been established. The few existing concrete efforts have assessed procedural fairness through checks on aspects such as the use of inappropriate or sensitive features (e.g., race, gender) during training, or by directly excluding these features \cite{10.1145/3461702.3462585,10.5555/3504035.3504042,10.1145/3593013.3594076}. 

\subsection{Stakeholder Involvement in AI Fairness Assessment}
Beyond AI experts and regulators, there are growing calls to involve stakeholders in AI fairness assessments \cite{10.1145/3631700.3664912,2024crowdsource/10.1145/3640543.3645209,10.1145/3630106.3659044,10.1145/3711079, 10.1145/3359283,10.1145/3514258,earnfairness, 2023/10.1145/3579601, 10.1145/3359284}. This has been partly in response to prior work which revealed that while experts rely on group-level operational metrics for statistical feasibility \cite{hort2023bias, 10.1145/3194770.3194776, 10.1145/3457607, conflict/10.1145/3351095.3372864}, stakeholders emphasize fine-grained, individual-level fairness, or more comprehensive notions that combine both individual- and group-level fairness \cite{earnfairness, doi:10.1080/10447318.2022.2067936}.

Most existing work focuses on one aspect: fairness metric preference. One approach collects instance-level feedback by asking stakeholders to judge or relabel individual AI predictions \cite{10.1145/3514258, 10.1145/3631700.3664912,taka2024human}. Another collects model-level feedback by having stakeholders pick AI models that result in preferred outcomes ~\cite{2024crowdsource/10.1145/3640543.3645209, MathematicalNotions/10.1145/3292500.3330664,Yokota2022Toward,Saxena/10.1145/3306618.3314248}. In both cases, fairness metric preferences are learned indirectly from these judgments. A different approach directly asks stakeholders to select from predefined fairness metrics, by explaining metrics to them directly with intuitive metric visualizations \cite{earnfairness, 10.1145/3411764.3445308,study3}. However, all approaches `limit' stakeholders' fairness values to existing metrics.

Recent studies show that stakeholders need and prefer custom fairness metrics \cite{study3, doi:10.1080/10447318.2022.2067936}. Work on creating fairness criteria has begun, but it is still in an early stage \cite{study3,10.1145/3630106.3659044,doi:10.1080/10447318.2022.2067936}, e.g., enabling end-users to customize the features they wish to consider, rather than being restricted to predefined protected features \cite{7fa9650f341548199b607745c62dc11c}, and working with stakeholders in workshop settings to express fairness concerns and co-design fairness notions \cite{10.1145/3630106.3659044}. Interactive tools have also gained attention, e.g., a proposed `fairness metric builder' that allows stakeholders to define their own mathematical formulas \cite{doi:10.1080/10447318.2022.2067936}, though it appears difficult for those without AI expertise and has not been thoroughly evaluated. Existing work does not yet uncover how stakeholders create quantifiable, operational criteria or what types emerge, limiting guidance for AI experts' criteria decisions and inclusive fairness assessment tool design. Our work addresses this knowledge gap.
 
\section{Methods}
\subsection{Context Setup: AI Scenario and Participant Recruitment}
We chose the credit rating scenario because it is familiar to most people, requires little domain knowledge, and has been shown to support individuals without AI expertise in reasoning about AI fairness \cite{earnfairness, doi:10.1080/10447318.2022.2067936, 10.1145/3514258}. We used the German Credit dataset, a widely used real-world benchmark \cite{dataset, earnfairness, 10.1145/3613904.3642627, pagano2023bias, de2025towards, 8843908}, which contains 1,000 instances with 20 demographic and financial features for a binary classification task predicting `Good Credit' or `Bad Credit'. Fairness criteria used by AI experts commonly focus on two protected features, Age and Gender, and one context-sensitive feature, Foreign Worker status \cite{FairnessDatasetSurvey}. We trained a widely used logistic regression model \cite{hort2023bias}. Prior work shows that the model transparency helps stakeholders understand AI and fairness easily \cite{10.1145/3514258, 10.1145/3411764.3445308, earnfairness}. Our model was trained over 1000 iterations and achieved an accuracy score of 0.76 over an 80-20 split of training and testing data, consistent with prior work \cite{earnfairness}. The model is fast enough to provide real-time fairness assessment feedback.

We recruited 18 participants (8 men and 10 women, self-identified) who were decision subjects without any AI training or professional experience, through public postings and social media channels of a university. Participants had varied education levels, diverse professional backgrounds, and represented a wide range of national backgrounds, reflecting diverse lived experiences. Participants' mean age was 31 (SD=8.06), aligning with the typical age of current loan applicants \cite{LoanUSA, LoanUk}, and participants had varied loan experience (12 with prior loan approval; 5 with no loan experience; 1 undisclosed). Full participant information is in Appendix Table~\ref{app:Participant Information}. All participants provided informed consent; privacy and confidentiality were ensured. Each participant received an Amazon voucher equivalent to US~\$20. This study was reviewed and approved by the University of Glasgow College of Science \& Engineering Ethics Committee as a low-risk study.

\subsection{Session Procedure: Supporting Stakeholders Creating Fairness Criteria}
\label{sec:procedure}
We piloted individual sessions and a group workshop to refine study materials and ensure sufficient time for participants to create fairness criteria, then conducted the user study. All sessions were identical for each participant, and on average, each participant spent around 95 minutes in total. Participants verbalized their reasoning throughout the study. 

\textbf{Preparation} Participants reviewed the participant information sheet, confirmed understanding of expected outputs (articulating fairness notions, creating fairness criteria), provided informed consent, and completed a demographic questionnaire. Researchers then introduced the AI scenario by presenting sample test instances with their features (along with a \textit{Feature Information Sheet} describing all features in plain language \cite{dataset}, Appendix \ref{app: feature information sheet}), ground-truth labels, and AI predictions. Then, basic AI concepts were explained in lay terms: how the AI model was trained, how it worked, and features' importance of the logistic regression model. Researchers introduced what legally protected features are. 

\textbf{Criteria Creation and assessment} \textbf{(1)} Firstly, participants were asked:``\textit{What does fairness mean to you, or how would you expect to assess whether the AI is fair based on your own fairness values?}'' and participants freely reflected on and articulated their fairness notions in their own language. \textbf{(2)} Next, participants were introduced to existing fairness metrics. Researchers explicitly clarified the intention of introducing existing metrics:``\textit{These metrics serve as illustrative examples that may inspire how you later quantify your notions so that we can measure fairness, rather than altering your original notions}''. Pilot feedback also suggested that introducing common metrics supported understanding without influencing criteria creation. Participants were given \textit{Metric Explanation Cards} (Appendix \ref{app: Metric Explanation Cards}) that provided visual explanations of six group fairness metrics (Demographic Parity, Equal Opportunity, Predictive Equality, Equalized Odds, Outcome Test, and Conditional Statistical Parity) and two individual fairness metrics (Counterfactual Fairness and Consistency). Researchers also objectively explained each metric and fairness thresholds (i.e., as the maximum acceptable level of unfairness, measuring from 0\% to 100\%, beyond which an AI model would be considered unfair), so participants could later set their own thresholds. \textbf{(3)} Then, once participants felt ready, they began quantifying their notions into criteria. Researchers first introduced how to use the prototype. Participants could then quantify their notions directly within the prototype when this was supported (see section \ref{sec: prototype}). When the prototype did not support their ideas, participants explained what they wanted to assess and how, either verbally or by sketching on paper, and researchers helped structure these ideas into concrete criteria with participants reviewing and confirming them. \textbf{(4)} Finally, participants viewed fairness assessment results based on their criteria. The prototype automatically computed supported criteria, while unsupported criteria were manually implemented through researchers' real-time coding.

After creating criteria, participants completed the NASA Task Load Index (NASA-TLX) questionnaire \cite{NasaTLX}, a widely used instrument for measuring perceived workload, and a post-questionnaire with 10 multiple-choice items designed to test their mental models of fairness-related concepts, inspired by \cite{10.1145/3613904.3642106}. 
          
\subsection{Prototype}
\label{sec: prototype}
Custom fairness design tools are under-explored and drawing on existing research \cite{earnfairness, doi:10.1080/10447318.2022.2067936, 10.1145/3194770.3194776, 10.5555/3504035.3504042}, we designed a four-module prototype with flexible, composable elements that help participants quantify their notions into criteria and operationalize fairness assessment in real time. We validated and refined it through a design workshop before the study; design rationale and details can be found in Appendix \ref{app:Prototype Extra Information}.

\begin{figure}[htbp]
\centering
\includegraphics[width=\linewidth]{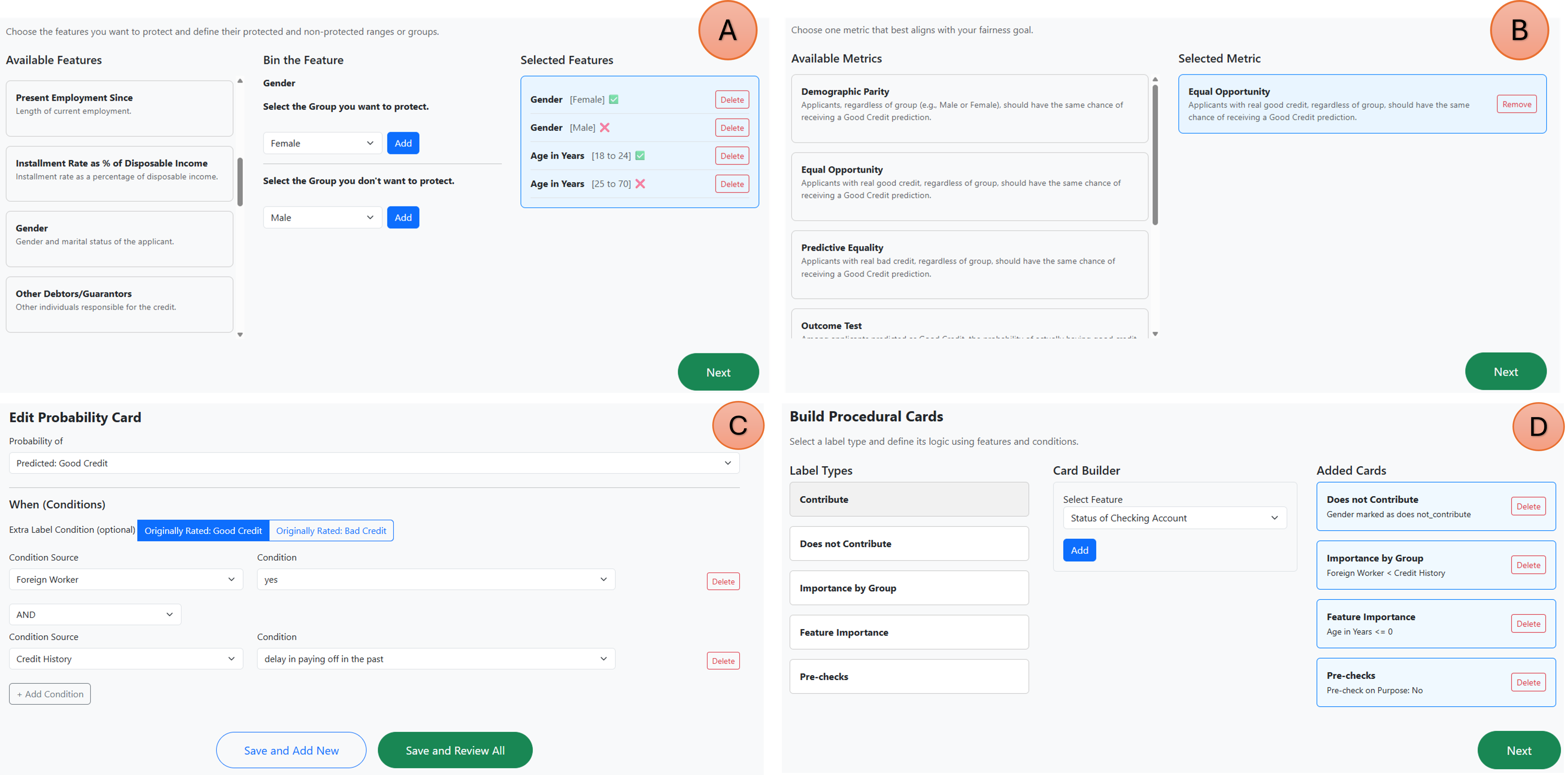}
\caption{Module 1: Using Existing Metrics. Participants selected features (A), e.g., Gender, binning Gender into protected (Female) and non-protected (Male), and chose existing fairness metrics (B), e.g., Equal Opportunity. Module 3: Creating new outcome fairness metrics (C). This figure shows an example metric defined by a participant: P(AI Predicted = Good Credit | Ground Truth =  Good Credit $\land$ Foreign Worker = Yes $\land$ Credit History = Delayed in paying off in the past). Module 4: Creating new procedural fairness metrics (D). This figure shows a procedural fairness criterion with 4 procedural rules: marking Gender as ``Does Not Contribute''; setting Foreign Worker to have lower importance than Credit History using ``Importance by Group''; setting the importance value of the Age feature to zero or lower by using ``Feature Importance''; and indicating ``no pre-checks'' for the feature Purpose.}
\label{fig:PrototypeOneImage}
\end{figure}

\textbf{Module 1 - Using existing outcome fairness metrics} (1) Feature Selection (Figure \ref{fig:PrototypeOneImage} A): ``Available Features'' lists all the features used for model training. Participants could select one or multiple features that were relevant to their fairness notions; ``Bin the Feature'' allowed participants to further bin specific feature values to indicate which groups they wanted to protect and which groups they do not want to protect based on their notions; ``Selected Features'' displays the features and their binning choices in real time, and participants could delete any selected feature or its binning to adjust at any time. (2) Metric Selection (Figure \ref{fig:PrototypeOneImage} B): ``Available Metrics'' lists all eight common existing metrics introduced to participants. Participants could click to select one metric that aligned with their fairness notion. ``Selected Metric'' displays the chosen metric in real time and allows participants to remove it and make a new selection. 
Participants then entered a fairness threshold value (from 0\%–100\%). Finally, the resulting criterion, showing the feature, metric, and threshold, was displayed for review, and participants could revise it.

\textbf{Module 2 - Combining multiple existing outcome fairness metrics}: Participants could create a fairness criterion by combining multiple outcome fairness metrics, with no limit on the metric number combined. Participants first used the same Module 1 (Figure \ref{fig:PrototypeOneImage} A and B) to set each sub-criterion. The prototype then supports several operators for combining each sub-criterion. Participants could select ``Boolean operators'' (e.g., ``AND'': all sub-criterion must be satisfied simultaneously, ``OR'': either one is satisfied), use``If-Not'' operator to set a priority order (e.g., apply one sub-criterion first, and only use the next if the first cannot be met), or input a ``Weight'' value (0–1) for each sub-criterion.

\textbf{Module 3 - Creating new outcome fairness metrics} (Figure \ref{fig:PrototypeOneImage} C): Prior research has shown that many common outcome fairness criteria are defined by mathematical relationships between predicted outcomes and ground-truth outcomes, often expressed as probability-based formulas \cite{10.1145/3194770.3194776}. Inspired by this, our system enabled participants to define new outcome fairness criteria by selecting and configuring probability relationships among three factors: ground-truth labels (which could be used to specify groups with a particular real outcome), AI-predicted labels (which could be used to specify groups with a particular prediction made by the AI model), and feature-based conditions (which could be used to specify groups that meet specific conditions by setting features with particular values). Figure \ref{fig:PrototypeOneImage} C shows how the module supported participants in constructing a custom formula. (1) The ``Probability of'' dropdown enabled participants to select the probability of interest, such as the probability of having a specific AI prediction category or a ground-truth category. (2) The ``When (Conditions)'' component further allowed participants to customize conditions by (i) setting an additional ground-truth label condition, and (ii) specifying feature-based conditions, where users select a feature (``condition source'') and define its corresponding value (``condition''). Multiple such conditions can then be combined using logical operators (e.g., ``AND'', ``OR''). 

\textbf{Module 4 - Creating new procedural fairness metrics }(Figure \ref{fig:PrototypeOneImage} D): Inspired by prior work that measures procedural fairness through the features used in decision making \cite{10.5555/3504035.3504042}, this module is designed to support participants in defining as many types of procedural rules as possible by: (1) Participants could select procedural rules according to their own fairness notions: ``Contribute'' means the feature's importance is greater than 0; ``Does Not Contribute'' means the feature's importance equals 0; ``Importance by Group'' allows participants to set whether the importance of one feature is greater than, less than, or equal to another feature; ``Feature Importance'' allows setting specific feature importance values for selected features; (2) ``Pre-Checks'' ensures procedural consistency: all individuals are processed using the same decision-making process, without exceptions for selected groups. (3) The ``Card Builder'' supports participants in applying these selected procedural rules to corresponding features. (4) The ``Added Cards'' section lists all the procedural rules participants have defined and the features to which they are applied. Participants could delete or modify any card at any time.

Details on real-time fairness assessment using participant-created criteria and thresholds are in Appendix \ref{app:Prototype Extra Information}.

\subsection{Data Collection and Analysis}
\label{sec: data analysis}
All sessions were audio-recorded, with screen recordings capturing participants' interactions with the prototype and researchers' real-time coding process. We also took pictures of any handwritten notes. We transcribed approximately 30 hours of audio using OpenAI's Whisper locally \cite{whisper}, and manually corrected the transcriptions for accuracy. First, we conducted a thematic analysis \cite{doi:10.1191/1478088706qp063oa} of the transcripts to identify how participants created their fairness criteria. Three researchers jointly developed an initial codebook, which two of them then independently applied to all 18 transcripts. Our team met weekly to resolve coding discrepancies, clarify definitions, and iteratively refine the codebook; after four iterations, it was finalized (Appendix~\ref{app:codebook}), with themes firmly grounded in the data. Second, to analyze what criteria were created, we used the prototype's automatic logs of criteria and combined them with those not supported by the prototype, drawing on participants' handwritten notes and researchers' real-time coding from screen recordings. We use descriptive statistics to quantify both the number of participants and the number of fairness criteria. We also collected NASA-TLX ratings to assess perceived task load and administered a 10-item post-questionnaire (10 points each, total = 100) to gauge participants' mental model of fairness-related concepts. 

\section{Results}

\subsection{RQ1: How do stakeholders create their own fairness criteria?}
Participants created fairness criteria in two phases: Grounding Fairness Notions Through Feature (Phase 1), where notions emerged, and Translating Notions into Operational Fairness Criteria (Phase 2), where notions were quantified and operationalized for assessment. Bold text denotes \textbf{themes}; `N' indicates the number of participants associated with each theme.

\subsubsection{Phase 1: Grounding Fairness Notions Through Features}
\label{sec: forming notions}
\begin{figure}[htbp]
    \centering
    \includegraphics[width=1\linewidth]{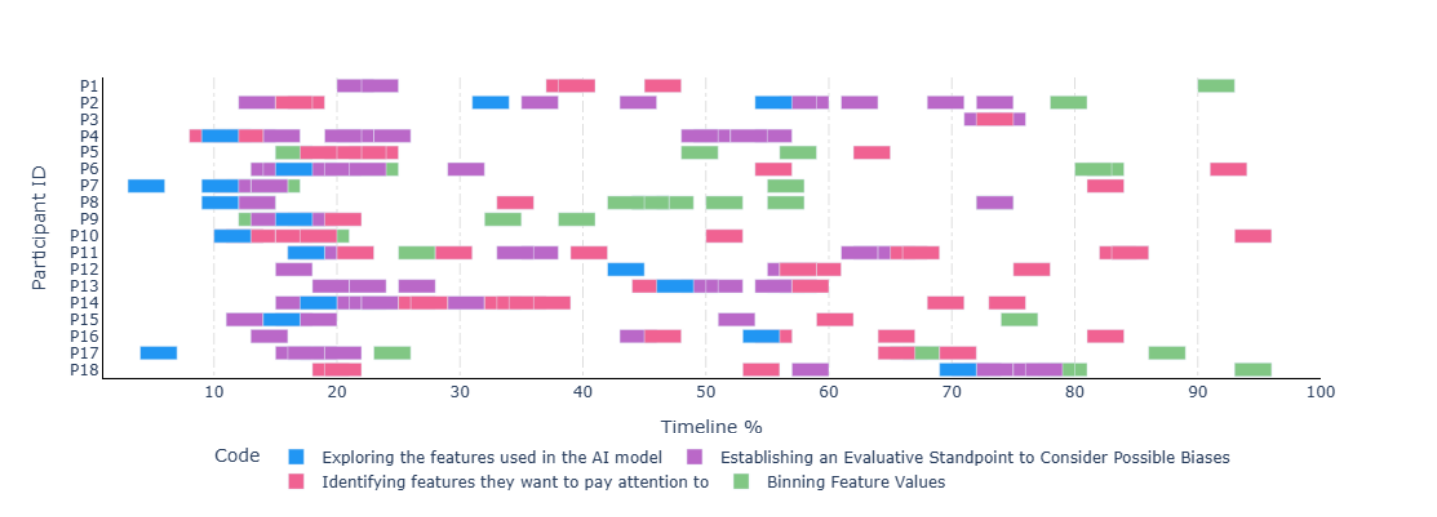}
    \vspace{-25pt}
    \caption{Actions with timeline for each participant in Phase 1: Grounding Fairness Notions Through Feature. The y-axis represents individual participants, with each row corresponding to one participant's creation process. Colored segments indicate themes associated with different actions involved in constructing fairness criteria. The x-axis represents the normalized progression of each participant's interaction session from start to finish (0–100\%), derived from each participant's session transcript using Nvivo. The position of each colored segment reflects when a particular action (i.e., theme) occurred within the overall creation process; the length of the segment represents the relative duration for which that action was present during the session.}
    \label{fig:phase1}
\end{figure}

Participants formed fairness notions by grounding them in model features through four actions (Figure \ref{fig:phase1}). Although not all participants engaged in all four actions and many iteratively moved between actions to clarify their notions, we identify a typical sequence of these actions across participants.
\vspace{-9pt}

Action 1: \textbf{Exploring the features used in the AI model} (N=15). Participants reviewed the feature set used for training the AI model and examined each feature's meaning, ensuring they understood what information the AI model relied on. Feature familiarization laid the groundwork for participants' subsequent fairness judgments.

Action 2: \textbf{Establishing an evaluative standpoint to consider possible biases} (N = 16). Participants demonstrated two reasoning approaches: a broader concern for objective fairness that benefits society beyond themselves, alongside reflection on their own demographics and desire for fair treatment. Most participants (N = 13) deliberately adopted an empathetic perspective, i.e., \textbf{Standing in someone else's shoes} to consider what kinds of biases the general public might face in this scenario rather than focusing only on biases that could affect themselves. Within this stance, the most popular way was to consider \textbf{Historical biases} (N = 6) based on common knowledge, such as gender discrimination or age-based prejudice. As P6 stated: \textit{`` I would worry that there would be a built-in bias against particularly women. ... Because of a historical bias against women, there could be a built-in bias against women when you're taking in those details''}. In contrast, a few participants (N = 5) voiced a self-referential perspective, \textbf{Projecting their own disadvantaged personal characteristics and experiences} (N = 5) to reason about potential biases. As P12 noted, \textit{``I'm an immigrant. I was rejected before for that reason''}, motivating P12 to protect against potential nationality or racial bias.

Action 3: \textbf{Identifying features they want to pay attention to} (N = 18). Half of the participants identified \textbf{Features that should not be used} (N = 9) in the AI model at all, believing that including these features in AI training would directly lead to unfairness. As P1 explained, \textit{``Why do you need to collect data such as gender, foreign worker or things like that? Wouldn't it just be fair to just not collect it at all? [...] because we're going to exclude them anyway. Unless you want to collect the data for like, I don't know, other reasons, like statistical reasons or something. Right after the fact, I feel like maybe just exclude all of that in the AI [training]''.} In contrast, participants also took a more lenient view: certain features could still be used in training the AI model, as long as these \textbf{Features should not contribute or have low importance} (N = 8) in the model's predictions, that is, their feature importance should be zero (not contribute) or close to zero (low importance).

Action 4: \textbf{Binning feature values} (N = 13). Finally, after participants had already identified features they wanted to pay attention to, participants actively redefine protective boundaries based on their situated understanding of vulnerability. For example, P7 binned Age into protected groups, under 25 and over 65, as populations that should not face age-based discrimination, unlike conventional AI fairness settings (e.g., credit rating typically protects only those under 25 \cite{FairnessDatasetSurvey}). 

Through these actions, participants reflected on their fairness notions and made them explicit. Among them, 12 participants developed procedural fairness notions centered on the use of features, which aligns with the idea that procedural fairness can be achieved by checking feature usage \cite{10.5555/3504035.3504042, 10.1145/3593013.3594076}. Fourteen participants formed outcome fairness notions, focusing on ensuring that the AI's prediction outcomes treat people fairly. Additionally, 8 participants incorporated both perspectives, emphasizing fairness in both the decision process and the prediction outcomes.

Participants who reasoned that certain features should not be used in the AI model naturally gravitated toward defining fairness through procedural fairness perspective, forming fairness notions around which features should be included or excluded to ensure fair decision-making processes. As P12 mentioned, fairness in AI decision-making is achieved when the system excludes irrelevant features Gender, Telephone, and Purpose, and only includes features indicative of an applicant's repayment ability: 
\vspace{-0.5em}
\begin{quote}
\textit{[I] would expect AI to make decisions based on my savings and salary, rather than my gender, telephone, and purpose, as I said before. [...] In this scenario, AI is more fair to everyone because they just includes all the information technically needed, because they can evaluate the people's ability to repay the loan. But sometimes some age groups indeed have a higher risk, so I can understand why the bank might [use it and] have a little bias on that. So I think it's the fairest way.}
\end{quote}
\vspace{-0.5em}

Participants who reasoned that certain features could be included in the AI model but should not contribute or should have low importance to its predictions defined fairness as procedural fairness, but instead engaged in specifying feature importance in the AI model or/and stipulating that certain features should not exceed others in importance (e.g., feature A should not be more important than feature B). As P4 illustrated: 
\vspace{-0.5em}
\begin{quote}
\textit{So basically what fairness means to me is [...] I think gender should be passed [to AI], but it shouldn't contribute [to AI's decision-making]. [...] I don't think that [Telephone] would have much impact, but it should be a part. [...Moreover,] while the model is learning, the importance of credit history should be less than importance of, let us say, checking account. And importance of credit history should be less than savings.}
\end{quote} 
\vspace{-0.5em}

Participants who attended to certain features formed from an outcome fairness perspective focused on measuring outcomes to ensure the AI model does not discriminate against specific people based on feature values. As P6 defined, gender might be present in the model but should not affect prediction outcomes: 
\vspace{-0.5em}
\begin{quote}
\textit{I would be more comfortable if you could pass on the gender characteristic [to AI]. [...] For two identical [individuals], there shouldn't be any prediction difference if it's just based on gender.}   
\end{quote}
\vspace{-0.5em}

\subsubsection{Phase 2: Translating Notions into Operational Fairness Criteria} 
\label{sec: translate notions into metrics}
In the second phase, participants turned fairness notions into concrete criteria by creating metrics, setting fairness thresholds and criteria priorities.

Action 1: Creating Metrics. All participants' first action was to concretize their criteria into a measurable form through metrics. We observed four distinct approaches to this action. The most common approach was that participants created \textbf{entirely new procedural fairness criteria} (N = 10) for measuring their procedural fairness notions. They focused on how AI decisions were made, primarily by specifying how features were used in the AI model. In doing so, they either selected and excluded features, setting the feature importance to indicate contributions for selected features, or ranked feature contributions. 

To quantify outcome fairness notions, participants adopted three approaches. The most common approach was directly building their metrics by \textbf{using one existing metric} (N = 9) when it already aligned with their fairness notions, selecting relevant features they had identified from Phase 1. For example, P3 initially expressed their notion as \textit{``as long as two or more people, they all have similar characteristics, they should be treated the same''} and then chose Metric-Consistency to represent this notion. However, using a single existing metric did not capture participants' fairness notions in some cases. Some participants created their criteria by \textbf{combining multiple existing metrics} (N = 7). Participants often began by choosing one existing metric that captured part of their fairness notion. They then identified what was missing and introduced additional ones, and finally decided how to combine them.

We observed that existing metrics alone did not capture participants' fairness notions. There were some participants \textbf{creating entirely new outcome metrics} (N = 6), by developing new mathematical formulas to evaluate the fairness of AI results. When creating these criteria, participants first identified which groups should be compared, that is, people from protected versus non-protected groups based on certain feature values, such as gender being female versus gender being male. They then refined these groups by adding more specific conditions. These conditions were defined by indicating people with certain ground-truth labels, model predictions, or other feature values. For example, a participant might refine these groups by filtering individuals whom the AI system predicted as having ``Good Credit'', that is, by setting the AI prediction label to Good Credit. Finally, they decided how to compare these groups in a quantified way to assess whether the AI system treated them equally.

Action 2: Setting Fairness Thresholds. Next, participants began \textbf{setting custom fairness threshold} (N = 18) to indicate how much unfairness they would like to accept (from 0\% to 100\%, as the deviation from the ideal assessment results based on the criteria they built), with most thresholds falling within 0–20\%, reflecting participants' high fairness expectations. Most participants (16 of 18) used a single fixed threshold, deeming a model unfair if the threshold was exceeded. However, two participants applied \textbf{multi-level thresholds}, rejecting the idea of using a fixed cutoff to decide when an AI system is fair, and setting more than one threshold for different situations, because they viewed fairness and thresholds as dynamic and context-dependent. As P3 highlighted, the fairness threshold depended on the overall credit score distribution and could be adjusted according to what a bank considers acceptable. 

Action 3: Setting Criteria Priorities. When defining multiple criteria, participants engaged in \textbf{setting the priority of criteria} (N = 13). Eight participants used an \textit{``x out of y must pass''} rule scheme, including two applied an \textit{``all-pass rule''}, requiring every criterion to hold for the AI system to be considered fair. In contrast, the remaining participants used a \textit{``(top-)k priority rule''}, requiring only the (top) K criteria to pass, considering practical constraints. 

\subsection{RQ2: What kinds of fairness criteria do stakeholders create?}
\label{sec:assessment_criteria}
\begin{figure}[H]
    \centering
    \includegraphics[width=0.9\linewidth]{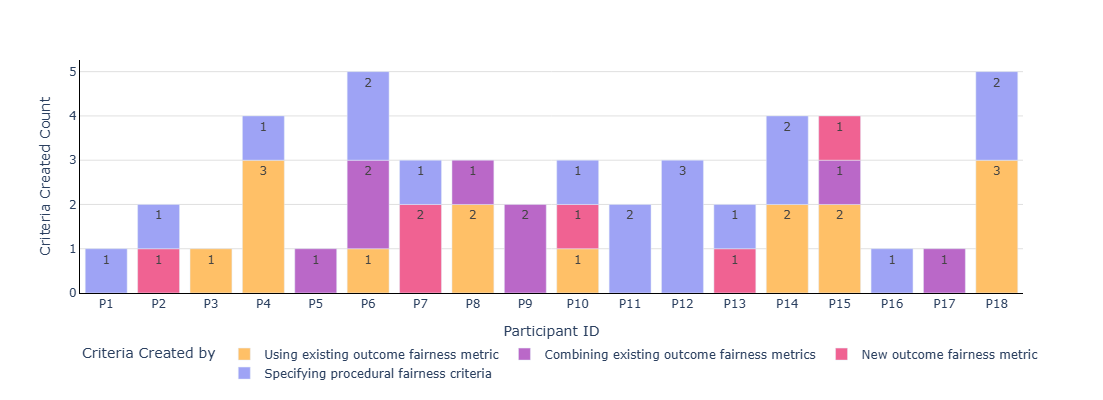}
    \vspace{-15pt}
    \caption{Different Ways of Criteria Created with Count by Participant}
    \label{fig:count_of_metrics}
\end{figure}

As the output of the previously described process, eighteen participants created a total of 47 fairness criteria, with a mean of 2.61 criteria per person (SD = 1.38; min = 1; max = 5). This suggests that participants often applied multiple criteria when assessing fairness, rather than the single-metric approaches widely adopted in current AI practice \cite{de2025towards,hort2023bias}. Of the 47 criteria (Figure~\ref{fig:count_of_metrics}), 15 criteria (8 participants) were directly using existing outcome fairness metrics. The remaining 32 criteria (17 participants) were created in three other ways: 8 criteria (6 participants) involved combining multiple existing outcome fairness metrics; 6 criteria (5 participants) involved creating new outcome fairness metrics; 18 criteria (12 participants) involved creating new procedural fairness metrics. Each created criterion was grounded in either an outcome fairness or a procedural fairness perspective, but some participants generated criteria across both, highlighting an intertwined understanding of fairness. Among the 18 participants, 14 participants created criteria for assessing outcome fairness, and 12 participants created criteria for assessing procedural fairness. Eight participants created criteria addressing both aspects. We now clarify the notation used and detail each criterion type.
\vspace{-12pt}
\vspace{-15pt}

\noindent
\colorbox{cyan!10}{
\parbox{\dimexpr\linewidth-2\fboxsep\relax}{
$P(\cdot)$ represents a (conditional) probability of relevant events.\\ 
$\omega$ represents feature importance; i.e., feature weight in our logistic regression model.\\ 
$1\{\cdot\}$ is an indicator function, returning 1 if the given condition is satisfied, otherwise returning 0.\\
$\land$ represents Boolean symbol AND. \\
$F_{\text{model}}$ denoted the feature set used in model training. \\
Criterion(PID\_n): the fairness criterion created by participant PID, with n showing how many criteria the participant created in total.
}}

\subsubsection{Using existing outcome fairness metrics}
Fifteen criteria (8 participants) were created by directly using one of the eight existing fairness metrics we provided to features they perceived as potential sources of bias. Only 2 criteria (created by 2 participants) focused on protected features, one on Age and one on Gender, while 13 criteria (6 participants) involved using unprotected features. This shows that stakeholders interpret potential bias more broadly, beyond legally protected features. For example, P8 created a fairness assessment criterion by selecting Equal Opportunity, applying it to the feature Job, and designating ``unskilled'' as the protected group, with other values (i.e., other job types) considered the non-protected group, ensuring that individuals with `Good Credit' have the same probability of being predicted as such across groups:

\noindent
\colorbox{cyan!10}{
\parbox{\dimexpr\linewidth-2\fboxsep\relax}{
 \textbf{Criterion(P8\_3): }\\
 P (Predicted=Good Credit | Ground Truth=Good Credit, Job=Unskilled) = P (Predicted=Good Credit | Ground Truth=Good Credit, Job=Others)
  }
}

\subsubsection{Combining existing outcome fairness metrics}
Criteria were also created by incorporating multiple existing outcome fairness metrics (8 criteria, 6 participants). All these criteria were created by using logical operators to combine these multiple existing metrics into one metric that could more fully represent their notion of fairness. The most common way was to use ``AND'' (8 criteria, 6 participants). No other Boolean operators were used, reflecting that participants wanted to enforce strict fairness rules, requiring all sub-metrics to hold at the same time.

Typically, participants only combined two metrics to meet their fairness criteria; only one participant (P17) created a criterion that combined three different metrics. Among them, the most frequent combination was Counterfactual Fairness and Demographic Parity (3 criteria, 2 participants), combining both individual and group outcome fairness metrics. For example, P6 used these two metrics to create a fairness criterion applied to Gender, to assess whether the AI model treated different genders fairly. The resulting criterion considered a model fair only when both of the following requirements were satisfied: for each individual, changing a person's gender did not affect the model's prediction (Counterfactual Fairness); and across gender groups, men and women had equal chances of receiving a ``Good Credit'' prediction from the AI model (Demographic Parity):

\noindent
\colorbox{cyan!10}{
\parbox{\dimexpr\linewidth-2\fboxsep\relax}{
 \(
\textbf{Criterion(P6\_5):} \\
\frac{1}{N} \sum_{(X, A_{\text{true}}) \in \text{dataset}} 1 \{p(X, A_{\text{female}}) = p(X, A_{\text{male}})\}
\)  $\land$ 
P(Predicted=Good Credit | Female) = P (Predicted=Good Credit | Male)
}}

\subsubsection{New outcome fairness metrics}
Some participants created new outcome fairness metrics (6 criteria, 5 participants) to capture nuanced and context-specific comparisons, which were more complex than existing metrics, such as Demographic Parity and Predictive Equality. We noted two patterns of constructing new outcome fairness metrics. The first pattern (3 criteria, 2 participants) was grounded in comparisons between participant-defined protected and non-protected groups. Each fairness criterion specified who should be compared, defined through feature–value pairs that indicated protected and non-protected groups. These groups were then refined by adding conditions, for instance, limiting the comparison to individuals defined by certain ground-truth labels, AI prediction outcomes, or other feature–value conditions. Finally, the criteria indicated how to measure equal treatment between these groups, often using mathematical relations to express fairness.

For example, P7 created a fairness criterion to ensure equal treatment for foreign and non-foreign workers, specifically among individuals whose ground truth was `Bad Credit' and whose loan Duration was within one year, and the AI model should give an equal chance of being predicted as Good Credit for both groups:

\noindent
\colorbox{cyan!10}{
\parbox{\dimexpr\linewidth-2\fboxsep\relax}{
 \textbf{Criterion(P7\_3): }\\
  P(Predicted=Good Credit | Ground-Truth=Bad Credit, Duration $\in$ {0-24}, Foreign Worker=Yes) = P(Predicted=Good Credit | Ground-Truth=Bad Credit, Duration $\in$ {0-24}, Foreign Worker=No)
  }
}

Interestingly, P7's criterion is an improvement upon the existing fairness metric, Predictive Equality. In Predictive Equality, fairness is achieved when, among people who actually have `Bad Credit', the probability of being predicted as `Good Credit' by the AI model is equal for foreign and non-foreign workers.  

The second pattern was aligning fairness with the AI model's overall predictive performance, rather than focusing on specific comparisons between protected and non-protected groups (3 criteria, 3 participants). All created criteria used the False Negative Rate (FNR), measuring the risk that individuals with Good Credit are incorrectly predicted as Bad Credit, which reveals a need for fairness that incorporates model error rates to better protect qualified individuals. Although FNR is commonly used by AI experts \cite{10.1145/3194770.3194776}, it was not provided and was independently derived by participants; we therefore treat it as a new outcome fairness criterion. For example,

\noindent
\colorbox{cyan!10}{
\parbox{\dimexpr\linewidth-2\fboxsep\relax}{
\text{\textbf{Criterion(P15\_4):}}
\[
\hspace*{-5cm}
\frac{\text{False Negatives (FN)}} {\text{True Positives (TP)} + \text{False Negatives (FN)}}
\]
}}

\subsubsection{Specifying procedural fairness criteria}
Most created criteria(18 criteria, 12 participants) focused on procedural fairness with participants defining different feature usage principles. More than half of these criteria (10 criteria, 9 participants) were about explicitly specifying that certain features should be excluded in model training in the first place. Excluding irrelevant features, this interpretation also resembles the idea of \textit{fairness through unawareness} \cite{10.5555/3504035.3504042, 10.1145/3457607}, where an algorithm is considered fair as long as ``protected features'' are not explicitly used in the decision-making process. However, it is worth noting that our participants not only focused on excluding protected features but also considered removing non-protected features that might introduce potential bias. For example, P11 specified that Gender, Foreign Worker, and Telephone (non-protected feature) must be excluded in training the model, as their presence in the dataset would make the AI system unfair under this criterion:

\noindent
\colorbox{cyan!10}{
  \parbox{\dimexpr\linewidth-2\fboxsep\relax}{
  \text{\textbf{Criterion(P11\_2): }} 
    \[
    \hspace*{-5.2cm}
    \mathbf{1}\{
    (\text{Gender} \notin F_{\text{model}}) \land
    (\text{Foreign Worker} \notin F_{\text{model}}) \land
    (\text{Telephone} \notin F_{\text{model}})
    \}
    \] 
  }
}

The remaining criteria (8 criteria, 8 participants) reflected more complex rules of feature usage: choosing which features to include, setting the feature importance values, and ranking features by feature contribution. For example, P16 created: a fair AI should focus on using financial factors like Credit History and Savings to make predictions (i.e., selecting features to include); demographic characteristics like Gender and Age could be included in the model, but should not contribute to making decisions (i.e., setting the feature importance values); and prioritize Credit history's influence on making predictions above all other features (i.e., ranking features by feature contribution). 

\noindent
\colorbox{cyan!10}{
\parbox{\dimexpr\linewidth-2\fboxsep\relax}{
\text{\textbf{Criterion(P16\_1):}}
\vspace{-6pt}
\[
\resizebox{\linewidth}{!}{$
\mathbf{1}\{
    (\omega_{\text{Credit History}} \neq 0) \land
    (\omega_{\text{Savings}} \neq 0) \land
    (\omega_{\text{Gender}} = 0) \land
    (\omega_{\text{Age}} = 0) \land
    (\omega_{\text{Credit History}} > \omega_f,~\forall f \in F_{\text{model}} \setminus \{\text{Credit History}\})
    \}
$}
\]
}
}

\section{Discussion} 
\subsection{Limitations and Future Work}
We acknowledge three limitations. First, our study examined only one domain (credit rating), limiting the insights on domains with different risk profiles (e.g., hiring, healthcare, criminal justice), where fairness criteria may differ due to context dependence. Further work is needed to inform diverse domain-specific mitigation and design. Second, our sample was limited and collected in person, which likely reduced the diversity of perspectives. Recent research calls \cite{10.1145/3630106.3659044,10.1145/3334480.3375158,doi:10.1080/10447318.2022.2067936,FairnessAssessmentPractionerPerspective,landers2023auditing} for involving stakeholders from a broad range of demographic backgrounds and especially marginalized groups. Future work could involve larger samples and examine how participants' backgrounds shape their criteria. Third, while our prototype supported both outcome and procedural fairness, it did not cover all possible criteria (e.g., causal reasoning \cite{10.1145/3194770.3194776}) at the design stage and did not allow participants to express two types of criteria: feature exclusion for procedural fairness and accuracy-based outcome criteria. Nonetheless, it was effective overall in supporting participants' criteria creation. Future work should extend tool support to enable more flexible and customized fairness design.

\subsection{Supporting Stakeholders' Fairness Criteria Creation}
Participants appeared to develop a solid understanding of AI fairness, averaging 88.9 out of 100 (SD=14.5) on the 10-item post-questionnaire. Despite no prior AI expertise and considerable `mental demand' (M=11.89) and `effort' (M=10.61), they reported low overall workload on NASA-TLX score (average 7.00 out of 21, SD=2.48, Min=2.33, Max=11.33) with high confidence in their `performance' (M=5.17), where lower scores indicate better outcomes (Appendix \ref{app:nasa}). 

\subsubsection{Process Design Implications: Supporting Fairness Reflection}
To reduce cognitive effort and enable broader participation, the process should first help stakeholders develop an understanding of fairness and reflect on their own notions, as we found some participants struggled to articulate what fairness means to them. Information that encourages reflection on potential biases in outcomes, procedures, and legally protected attributes should be provided, while remaining neutral and avoiding steering stakeholders' thoughts. Support should also encourage perspectives beyond personal experience, including societal and historical biases, as many participants demonstrated. Meanwhile, self-focused biases such as outcome favorability bias, where receiving favorable outcomes (e.g., ``Good Credit'' in credit rating) could increase people's perceived fairness, should be mitigated \cite{10.1145/3313831.3376813}. This can be mitigated by avoiding assigning personally favorable or unfavorable outcomes, as in our study. 

\subsubsection{Tool Design Implications: Supporting Stakeholder Creating Fairness Criteria}
Tool support for stakeholder-defined fairness criteria remains limited and difficult for non-experts \cite{doi:10.1080/10447318.2022.2067936}; building on this gap, we derive tool design implications from our prototype and observed criteria-creation processes for more inclusive fairness assessment.

\textbf{Providing Feature information and Model transparency} When forming fairness notions, participants relied on understanding model features and their values for outcome fairness criteria; procedural fairness criteria were also grounded in the perceived importance of these features in the trained AI model, highlighting the need to provide sufficient feature information. While feature importance was transparent in our logistic regression model via feature weights, more complex models require Explainable AI (XAI) techniques to improve transparency \cite{10.1145/3514258, IUI2021DesignMF}, such as SHAP for visualizing feature contributions \cite{10.1145/3324884.3418932}. Prior work has leveraged XAI to support fairness decisions, including FEE \cite{earnfairness}, Nakao et al.’s explanatory debugging system \cite{10.1145/3514258,5635185}, and FairHIL, which further incorporates causal graphs \cite{doi:10.1080/10447318.2022.2067936}. Future tools could also integrate counterfactual explanations to help stakeholders reason about how feature changes affect outcomes.

\textbf{Enable Stakeholders to Define Who to Protect} 
We found that 13 of 18 participants defined who to protect beyond legally protected attributes, incorporating contextually sensitive features and specific values, consistent with prior work showing that unprotected features shape fairness judgments \cite{study3}. For instance, some participants considered individuals with `Number of Dependents > 2' as a protected group. However, existing tools often restrict stakeholders to legally protected features \cite{10.1145/3411764.3445308, earnfairness, study3}. Tools should therefore allow stakeholders to define protected groups using any relevant features and support multi-feature subgroup definitions to capture intersectional biases \cite{8986948, earnfairness}.

\textbf{Enable Stakeholders to Define How to Measure Diverse Fairness Perspectives} Stakeholders should be supported in both understanding existing metrics and customizing their own criteria, as participants strongly preferred self-defined criteria (32 of 47). Through combining existing metrics, using Boolean operators, and using relationships between ground-truth labels, prediction outcomes, and complex feature conditions, participants were able to express the majority of their outcome fairness notions. While most fairness tools focus on outcome fairness \cite{doi:10.1080/10447318.2022.2067936, earnfairness, 10.1145/3411764.3445308, 2024crowdsource/10.1145/3640543.3645209}, we found that the majority of participants (12/18) valued procedural fairness, measuring how decisions are made rather than outcomes for fairness. Tools should enable stakeholders to actively adjust how features are used in the modeling process, for example, allowing them to decide which features to use for model training \cite{2023/10.1145/3579601}, to modify feature importance or weights \cite{10.1145/3514258, taka2024human, 10.1145/3631700.3664912}, or incorporating causal graphs for adjusting feature relationships \cite{10.1145/3194770.3194776}.

\textbf{Support Flexible Creation} As we described in Section \ref{sec: forming notions}, stakeholders frequently moved back and forth when defining their fairness notions, reflecting a non-linear and iterative process in formulating what fairness meant to them in context. Tools should therefore support this flexibility, allowing stakeholders to switch between features, notions, and criteria, and to edit and validate their ideas iteratively without being disrupted or overwhelmed. This is a challenging task, which calls for easy access to multiple information sources and seamless switching between actions. 

\subsection{\rev{Stakeholder-Created Fairness Criteria and Their Integration in Practice}}
Our findings show that features play a central role in how stakeholders create fairness criteria in predictive AI. Through their feature choices, stakeholders make explicit who they want to protect and how they believe fairness should be upheld. Rather than relying on a single existing metric (only one participant did so, Figure \ref{fig:count_of_metrics}), participants combined multiple metrics, especially combining group and individual fairness, indicating a need for more nuanced notions of fairness consistent with prior work \cite{earnfairness,study3,10.1145/3757647}. Beyond these studies, we show that stakeholders introduced new outcome metrics beyond predicted and ground-truth labels to capture context-specific concerns tied to valued features. Importantly, stakeholders also emphasized decision-process fairness by explicitly defining what and how features should be used, extending prior outcome-focused fairness.

The criteria reflect AI experts' view that single fairness metrics are insufficient, though guidance on combining metrics is lacking. Our findings might offer insights into combining metrics based on stakeholders' fairness needs for social inclusivity, such as pairing group and individual fairness rather than using group fairness alone, or further incorporating procedural fairness through feature usage measurement. Instead of requiring all criteria be met simultaneously ('AND'), weighted multi-objective approaches (with assigned metric weights) \cite{10.1145/3375627.3375862} or constraint hierarchies (with metric prioritization) \cite{earnfairness} could also be used, given that integrating these criteria may increase complexity as some metrics can conflict \cite{10.1145/3457607,conflict/10.1145/3351095.3372864,hort2023bias}.

\section{Conclusion}
We conducted a user study with 18 participants in a credit rating scenario to investigate both the process by which participants created their criteria and the fairness criteria they created. We observed that participants engaged in four actions to gradually clarify their fairness notions: exploring model features, considering potential biases, identifying features related to those biases, and refining protected people through feature binning. They then translated these notions into concrete criteria, and most participants created more than one criterion. They adapted existing outcome metrics, combined multiple metrics, or created totally new ones to assess outcome fairness. Notably, participants also concentrated on procedural fairness, expressing it through criteria that specified feature exclusion or feature importance in the model. Our work shows that involving stakeholders directly in this lengthy, demanding, and complex process is feasible, if the right information and the right tools are available to them, thereby ultimately paving the way toward more inclusive and responsible AI systems.

\begin{acks}
  We thank all participants for their contributions. We gratefully acknowledge the funding provided by the University of Glasgow and Fujitsu Limited. This work was also partially supported by the Engineering and Physical Sciences Research Council [grant number EP/Y009800/1], through funding from Responsible Ai UK (KP0011). 
\end{acks}

\bibliographystyle{ACM-Reference-Format}
\bibliography{references}

\clearpage

\appendix

\section{GenAI Usage Disclosure}
We hereby acknowledge that no generative AI tools (e.g., ChatGPT or other large language models) were used at any stage of this research, including during writing, data analysis, or code development. All content presented in this manuscript is the original work of the authors.

\section{Appendix}

\subsection{Participant Information}
\label{app:Participant Information}

Participants had different education levels (1 below high school level, 3 high school, 3 Bachelor's, and 11 Master's degrees), professional areas (e.g., law, history, education, healthcare, IT, and business), and nationalities (across Europe, Asia, Africa, and Latin America).

\begin{table}[h]
\centering
\resizebox{\textwidth}{!}{\begin{tabular}{lllllll}
\hline
PID & Gender & Age & Loan Experience & Area & Highest Education & Nationality \\
\hline
P1 & Female & 27 & Yes & History & Masters & British \\
P2 & Male & 25 & Yes & Education & Bachelors & British \\
P3 & Female & 40 & Prefer not to disclose & Healthcare and Medical & Below High School Education & Chinese \\
P4 & Male & 26 & Yes & Healthcare and Medical & Masters & Indian \\
P5 & Female & 25 & No & Healthcare and Medical & Bachelors & Chinese \\
P6 & Female & 26 & No & Education & Masters & Italian \\
P7 & Male & 27 & Yes & Firmware and Hardware Manufacturing & Masters & Indian \\
P8 & Female & 46 & No & Computer and IT (Managerial) & Masters & Irish \\
P9 & Female & 29 & No & Computer and IT & Bachelors & Iranian \\
P10 & Male & 20 & Yes & Geography & High School Education & Polish \\
P11 & Female & 30 & No & Law & Masters & Chinese \\
P12 & Female & 33 & Yes & Education & Masters & Chinese \\
P13 & Male & 38 & Yes & Education & Masters & Scottish \\  
P14 & Female & 45 & Yes & Business and Finance & Masters & Indonesian \\
P15 & Male & 41 & Yes & IT & High School Education & Salvadorean \\
P16 & Male & 19 & Yes & Computer and IT & High School Education & British \\
P17 & Male & 27 & Yes & Computer and IT & Masters & African \\
P18 & Female & 34 & Yes  & Healthcare and Medical (Managerial) & Masters & British\\

\hline
\end{tabular}}
\caption{Participant Summary} 
\end{table}
\clearpage

\subsection{Supporting Materials}
\label{app: Supporting Materials}
\subsubsection{Feature Information}
\label{app: feature information sheet}
This table presents the 20 features from the German Credit dataset that participants could refer to.
\begin{table}[h]
\centering
\resizebox{\textwidth}{!}{\begin{tabular}{llll}
\hline

Feature Name & Feature Description & Type & Example Value \\ \hline

Duration & Duration of the credit in months & Numerical & 30 (Months) \\
Credit Amount & Credit amount asked by the applicant & Numerical & 5234 (DM) \\
Installment Rate & installment rate in percentage of disposable income & Numerical & 4 \\
Residence Length & Length of time (in years) the applicant has lived in the present residence & Numerical & 1 \\
Existing Credits & Number of existing credits at this bank & Numerical & 1 \\
Dependents & Number of people being liable to provide maintenance for & Numerical & 1 \\
Age & The age of the applicant in years & Numerical & 28 \\
Gender & The gender of the applicant & Categorical & Male \\
Checking Account & Status/balance of checking account at this bank & Categorical & 0-200 DM \\
Credit History & Past credit history of applicant at this bank & Categorical & Other Existing Credits \\
Purpose & The applicant's purpose for the loan (e.g., car, education) & Categorical & Car (New) \\
Savings & Savings accounts/bonds at this bank & Categorical & <100 DM \\
Employment & Present employment since & Categorical & Unemployed \\
Debtors & Other debtors/guarantors present & Categorical & None \\
Property & Properties that applicant has & Categorical & Car/Other \\
Installment Plans & Other installment plans the applicant is paying & Categorical & None \\
Housing & Housing (rent, own, for free) & Categorical & Own \\
Job & Current job information (unemployed, (un)skilled, management) & Categorical & Management/Officer \\
Telephone & Is there any telephone registered for this applicant? & Binary & Yes \\
Foreign Worker & Is applicant foreign worker? & Binary & Yes \\

\hline

\end{tabular}}
\caption{German Credit Dataset: Features with their Descriptions}
\label{tab:german_features}
\end{table}

\subsubsection{Metric Explanation Card}
\label{app: Metric Explanation Cards}
\begin{figure}[h]
    \centering
    \includegraphics[width=0.98\linewidth]{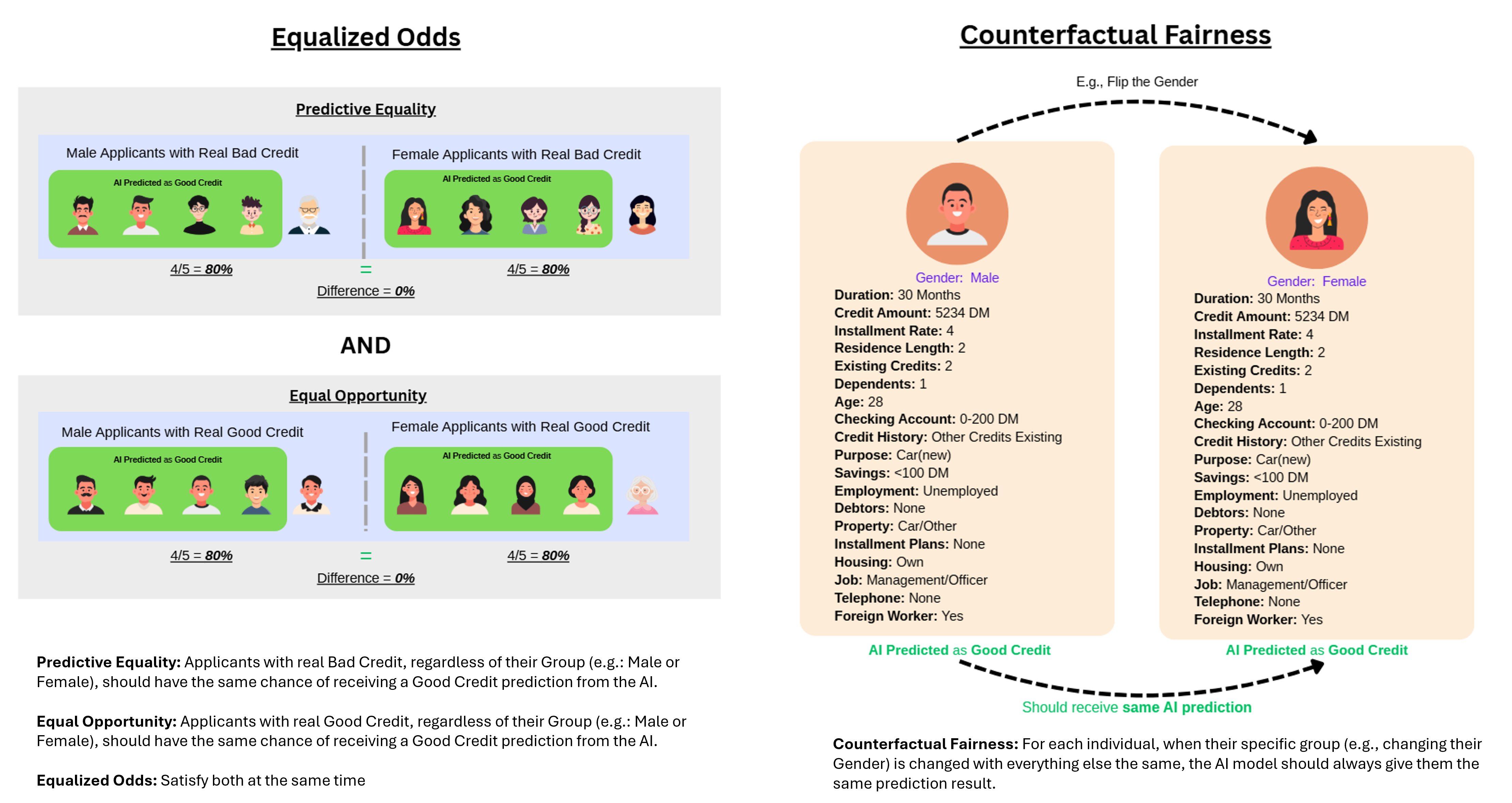}
    \caption{Metric Explanation Card: This figure illustrates the visualization and textual explanation of the Equalized Odds and Counterfactual Fairness metrics. Equalized Odds (composed of Equal Opportunity and Predictive Equality) represents a group fairness metric, while Counterfactual Fairness represents an individual fairness metric. All eight existing metrics follow a consistent explanation and visualization style.}
    \label{fig:metric}
    \end{figure}

Figure \ref{fig:metric} presents example metric explanation cards for Demographic Parity and Counterfactual Fairness.

\subsubsection{Prototype}   
\label{app:Prototype Extra Information}

\textbf{Design Rationale and User Testing} Considering that fairness is complex, that we do not yet have a good understanding of how stakeholders create fairness criteria, and that there are no mature interactive tools to draw upon that allow criteria creation, we never intended the prototype to be covering all criteria or to constrain stakeholders' creation process. Participants were free to use it or not; if not, researchers still facilitated their criteria formulation as described in Section~\ref{sec:procedure}.

Our design started out by drawing on prior research \cite{earnfairness, doi:10.1080/10447318.2022.2067936, 10.1145/3194770.3194776, 10.5555/3504035.3504042} and current practices (Section \ref{sec: AI fairness assessment practice}), which inspired the development of the four criteria-creation modules in our prototype: (1) using existing outcome metrics, (2) combining existing outcome metrics with Boolean logic, (3) creating new outcome fairness metrics, and (4) creating new procedural fairness metrics. 

Next, we created sketches for each of the functional modules, which we evaluated in a design workshop conducted prior to the main study with six participants without any AI background. These participants were distinct from the 18 participants in the main study. \rev{During the workshop, participants worked with the same scenario as in the main study. They first shared their own fairness notions and then tried out all parts of the sketched interface together, providing feedback on how the design could better help turn lay-language notions into concrete fairness criteria. We directly used this feedback to refine the prototype:} we reduced cognitive burden by splitting the creation of a criterion into smaller, switchable components over several screens; improved flexibility by supporting iterative criteria editing and naming of criteria; provided immediate feedback by supporting real-time assessment results of criteria against the model to ensure users were able to validate whether their criteria were functioning correctly and support incremental adjustments; improved usability by adopting clickable interface components; enabled deeper criteria creation by providing richer logical operators for combining existing metrics when constructing criteria.

\textbf{Real-Time Fairness Assessment Based on Created Criteria} The fairness threshold set by participants represented the maximum level of unfairness they considered acceptable, i.e., the largest tolerable deviation from the ideal result of their fairness criterion. In addition to supporting the creation of fairness criteria, the prototype also provided a calculation of the results of applying a criterion to the model, and showed the fairness outcome given the threshold that participants selected. To do so, criteria results based on outcome fairness were computed directly from the corresponding formulas. Because there is no equivalent process for procedural fairness, we took the decision to calculate the result as the percentage of participant-defined procedural rules that the model successfully followed. For instance, if a participant created five such rules for different features and the model complied with three of them, the procedural fairness score would be 60\%. Accordingly, the prototype classified a model as fair when its assessment result deviated from the ideal result by less than the specified threshold, and unfair otherwise.

Since the dataset occasionally lacks certain entries (e.g., when a participant-defined subgroup does not exist due to sparsity, making it impossible to assess fairness in real time for stakeholders), we used the Gaussian Copula Synthesizer from the Synthetic Data Vault (SDV) library \cite{SDV} to generate additional instances that preserve realistic feature dependencies.
\clearpage

\subsection{Codebook}
\label{app:codebook}
\begingroup
\small

\begin{longtable}{>{\raggedright\arraybackslash}p{2cm}
                  >{\raggedright\arraybackslash}p{3.5cm}
                  >{\raggedright\arraybackslash}p{7cm}
                  >{\raggedright\arraybackslash}p{0.6cm}
                  >{\raggedright\arraybackslash}p{0.6cm}
                  >{\raggedright\arraybackslash}p{0.6cm}}
\caption{Codebook \label{tab:themes}}\\
\toprule
\textbf{Theme} & \textbf{Code} & \textbf{Coded Text} & \textbf{PID} & \textbf{Files} & \textbf{Ref.} \\
\midrule
\endfirsthead

\multicolumn{6}{c}{{Codebook -- continued from previous page}} \\
\toprule
\textbf{Theme} & \textbf{Code} & \textbf{Coded Text} & \textbf{PID} & \textbf{Files} & \textbf{Ref.} \\
\midrule
\endhead

\midrule \multicolumn{6}{r}{{Continued on next page}} \\
\endfoot

\bottomrule
\endlastfoot

\textbf{Process} & & & & \textbf{18} & \textbf{529} \\
 & \multicolumn{3}{c}{\textbf{\textcolor{blue}{Phase 1: Grounding Fairness Notions Through Features}}} & \textbf{\textcolor{blue}{18}} & \textbf{\textcolor{blue}{326}} \\

 & Exploring the features used in the AI model & Gender seems like a pretty obvious one. Age as well. ... That seems like it could easily end up discriminating against people in 'unskilled' jobs. I personally don't think there is such a thing as an unskilled worker,but you see the larger point I'm making & P6 & 15 & 17 \\
 
\cdashline{2-6}[.8pt/2pt]

 & Establishing an evaluative standpoint to consider possible biases & Because I think like dependents ... making decisions irrespective of personal characteristics and irrespective of like things like that is kind of what would seem to be fair & P1 & 16 & 84 \\

\cdashline{2-6}[.4pt/4pt]

 & \textit{-- Standing in someone else's shoes} & So what is fair for me also is your previous experience. ... Fair for me is not just checking if I had an issue before,it's more about like, have I paid or not? & P15 & 13 & 45 \\

\cdashline{2-6}[.4pt/4pt]

 & \textit{-- Historical biases} & I would worry that there would be a built-in bias against particularly women. ... Because of a historical bias against women, there could be a built-in bias against women when you're taking in those details & P6 & 6 & 22 \\

\cdashline{2-6}[.4pt/4pt]

 & \textit{-- Projecting their own disadvantaged personal characteristics and experiences} & I'm an immigrant. I was rejected before for that reason & P4 & 5 & 9 \\

\cdashline{2-6}[.8pt/2pt]

 & Identifying features they want to pay attention to & For me, a fair,like an actual fair assessment,would be done pretty exclusively on the actual sort of financial history that I'm bringing in, and not on personal characteristics. & P6 & 18 & 59 \\

\cdashline{2-6}[.4pt/4pt]

 & \textit{-- Features that should not be used} & Why do you need to collect data such as gender, foreign worker or things like that? Wouldn't it just be fair to just not collect it at all?  [...] because we're going to exclude them anyway. Unless you want to collect the data for like, I don't know, other reasons, like statistical reasons or something. Right after the fact, I feel like maybe just exclude all of that in the AI [training] & P1 & 9 & 26 \\

\cdashline{2-6}[.4pt/4pt]

 & \textit{-- Features should not contribute or have low importance} & I don't think that [Telephone] would have much impact, but it should be apart [...Moreover,] while the model is learning, the importance of credit history should be less than importance of, let us say, checking account. And importance of credit history should be less than savings & P12 & 8 & 21 \\

\cdashline{2-6}[.8pt/2pt]

 & Binning feature values & the usual groups discriminated against are 18 to 25, maybe up to 30 & P6 & 13 & 43 \\

\cdashline{2-6}[.8pt/2pt]

& \multicolumn{3}{c}{\textbf{\textcolor{blue}{Phase 2: Translating notions into operational fairness criteria}}} & \textbf{\textcolor{blue}{14}} & \textbf{\textcolor{blue}{203}} \\

 & entirely new procedural fairness criteria & I would say that credit history is more important than purpose & P16 & 10 & 12 \\

 \cdashline{2-6}[.8pt/2pt]

 & using one existing metric & as long as two or more people, they all have similar characteristics, they should be treated the same & P3 & 9 & 22 \\

 \cdashline{2-6}[.8pt/2pt]

 & combining multiple existing metrics & An ideal model would satisfy all of these. ... For something to be totally fair, it feels like it should satisfy all of them. & P6 & 7 & 11 \\

 \cdashline{2-6}[.8pt/2pt]

 & creating entirely new outcome metrics & If there are two people, both have exactly the same things. One person applied for business loan, one person applied for TV, the person applying for business loan should be given priority & P9 & 6 & 14 \\

 \cdashline{2-6}[.8pt/2pt]

 & setting custom fairness threshold & because we are building my own idealized thing, I would like to put it down at zero. I realize that's probably not likely in practice, but I think that's what we should be striving for & P6 & 18 & 47 \\

\cdashline{2-6}[.4pt/4pt]

 & \textit{-- multi-level thresholds} & If out of all the conditions, if 80 to 100 percent of them are matched, then it's fair. If it is 70 to 79 percent match, it is still okay,partially fair. ... If it's less than 59, then it's unfair. & P3 & 2 & 4 \\

 \cdashline{2-6}[.8pt/2pt]

 & setting the priority of criteria & Okay, I think I'd weight the procedural one above the accuracy one because having it be fair in terms of it's not discriminating against gender or country of origin is more important ... than having it match tellers who may or may not be implementing those biases in the first place & P2 & 13 & 93 \\

\end{longtable}

\endgroup

\subsection{NASA-TLX}
\label{app:nasa}
To assess the cognitive effort involved in designing fairness criteria, we administered the NASA Task Load Index (NASA-TLX). Each dimension was scored on a scale from 0 to 21, and for all measures (including `performance', which we reversed), lower scores indicate higher confidence in performance. Table \ref{tab:nasa_tlx} reports each participant's scores on all dimensions, along with the averages.

\begin{table}[ht]
\centering
\caption{NASA-TLX Participant-Wise Scores}
\label{tab:nasa_tlx}
\begin{tabular}{llllllll}
\toprule
 PID & Metal Demand & Physical Demand & Temporal Demand & Performance & Effort & Frustration & NASA\_TXL Raw Score \\
\midrule
 P1  & 5  & 1 & 5  & 7 & 10 & 6  & 5.67 \\
 P2  & 16 & 1 & 1  & 4 & 15 & 4  & 6.83 \\
 P3  & 5  & 5 & 5  & 4 & 5  & 5  & 4.83 \\
 P4  & 8  & 3 & 8  & 7 & 5  & 1  & 5.33 \\
 P5  & 15 & 5 & 13 & 4 & 18 & 3  & 9.67 \\
 P6  & 12 & 1 & 9  & 1 & 14 & 1  & 6.33 \\
 P7  & 5  & 1 & 1  & 1 & 5  & 1  & 2.33 \\
 P8  & 15 & 1 & 5  & 17  & 15 & 5  & 9.67 \\
 P9  & 5  & 1 & 1  & 1 & 1  & 5  & 2.33 \\
 P10 & 11 & 1 & 2  & 2 & 13 & 7  & 6.00 \\
 P11 & 17 & 5 & 10 & 2 & 10 & 10 & 9.00 \\
 P12 & 20 & 5 & 1  & 1 & 21 & 1  & 8.17 \\
 P13 & 14 & 1 & 15 & 17  & 9  & 12 & 11.33 \\
 P14 & 18 & 5 & 8  & 2 & 17 & 15 & 10.83 \\
 P15 & 15 & 5 & 4  & 12 & 10 & 19 & 10.83 \\
 P16 & 10 & 2 & 2  & 7 & 5  & 3  & 4.83 \\
 P17 & 13 & 1 & 9  & 3 & 13 & 2  & 6.83 \\
 P18 & 10 & 5 & 5  & 1 & 5  & 5  & 5.17 \\
 \cdashline{1-8}[.6pt/2pt]
 \textit{Average} & \textit{11.89} & \textit{2.72} & \textit{5.78} & \textit{5.17} & \textit{10.61} & \textit{5.83} & \textbf{7.00} \\
\bottomrule
\end{tabular}
\end{table}

\end{document}